\documentclass[amsmath,amssymb,aps,prl,twocolumn,notitlepage,superscriptaddress,longbibliography]{revtex4-2}

\usepackage{bm}
\usepackage{amsmath}
\usepackage{amssymb,amsthm}
\usepackage{mathrsfs}
\usepackage{bbm}
\usepackage{epsfig,color}
\usepackage[sans]{dsfont}
\usepackage{comment}
\PassOptionsToPackage{hyphens}{url}
\usepackage{hyperref}
\usepackage{url}
\usepackage{units}
\usepackage{color}
\usepackage{ifthen}
\usepackage{graphicx}
\graphicspath{ {graphics/} }
\usepackage{tikz-cd}
\usepackage[utf8]{inputenc}
\usepackage{tikz}
\usetikzlibrary{positioning}
\usetikzlibrary{automata,positioning}
\usetikzlibrary{fit, shapes.geometric}
\usetikzlibrary{decorations.pathmorphing}
\usetikzlibrary{arrows,matrix}
\usepackage{tikz-3dplot}
\usetikzlibrary{patterns}
\usetikzlibrary{shapes.geometric}
\usepackage[normalem]{ulem}

\usepackage{mathtools}
\DeclarePairedDelimiter{\abs}{\lvert}{\rvert}
\definecolor{nblue}{rgb}{0.2,0.2,0.7}
\definecolor{ngreen}{rgb}{0.2,0.6,0.2}
\definecolor{nred}{rgb}{0.7,0.2,0.2}
\definecolor{nblack}{rgb}{0,0,0}

\newcommand{\ket}[1]{|#1\rangle}
\newcommand{\pasm}{process causal asymmetry}

\newcommand{\ketbra}[2]{\ket{#1}\!\bra{#2}}
\newcommand{\bra}[1]{\langle#1|}

\newcommand{\tr}{\text{tr}}

\def\A{\mathcal{I}}
\def\B{\mathcal{B}}

\newtheorem{result}{Result}

\def\A{\mathcal{A}}
\def\B{\mathcal{B}}

\def\S{\mathcal{S}}

\def\T{\mathcal{T}}
\def\X{\mathcal{X}}
\def\Y{\mathcal{Y}}

\newcommand{\braket}[2]{\langle #1 \vert #2 \rangle}

\newcommand{\ema}{$\epsilon$-machine}
\newcommand{\emas}{$\epsilon$-machines}
\newcommand{\etr}{$\epsilon$-transducer}
\newcommand{\etrs}{$\epsilon$-transducers}

\newcommand{\er}{\sim_\epsilon}

\def\tr{\mbox{tr}}

\def\bea{\begin{eqnarray}}
	\def\eea{\end{eqnarray}}

\newcommand{\norm}[1]{\left\lVert#1\right\rVert}

\theoremstyle{definition}
\providecommand{\definitionname}{Definition}

\begin{document}

\title{Causal Asymmetry of Classical and Quantum Autonomous Agents}

\author{Spiros Kechrimparis}
\email{skechrimparis@gmail.com}
\affiliation{School of Computational Sciences, Korea Institute for Advanced Study, Seoul 02455, South Korea}
\affiliation{Nanyang Quantum Hub, School of Physical and Mathematical Sciences, Nanyang Technological University, 637371, Singapore.}

\author{Mile Gu}
\email{mgu@quantumcomplexity.org}
\affiliation{Nanyang Quantum Hub, School of Physical and Mathematical Sciences, Nanyang Technological University, 637371, Singapore.}
\affiliation{Centre for Quantum Technologies, National University of Singapore, 3 Science Drive 2, 117543, Singapore.}
\affiliation{MajuLab, CNRS-UNS-NUS-NTU International Joint Research Unit, UMI 3654, 117543, Singapore.}

\author{Hyukjoon Kwon}
\email{hjkwon@kias.re.kr}
\affiliation{School of Computational Sciences, Korea Institute for Advanced Study, Seoul 02455, South Korea}
	
\begin{abstract}  Why is it that a ticking clock typically becomes less accurate when subject to outside noise but rarely the reverse? Here, we formalize this phenomenon by introducing \emph{process causal asymmetry} -- a fundamental difference in the amount of past information an autonomous agent must track to transform one stochastic process to another over an agent that transforms in the opposite direction. We then illustrate that this asymmetry can paradoxically be reversed when agents possess a quantum memory. Thus, the spontaneous direction in which processes get `simpler' may be different, depending on whether quantum information processing is allowed or not.
\end{abstract}
\maketitle 

\begin{figure}[!t]
    \centering
    \scalebox{0.7}{
    \begin{tikzpicture}[node distance={20mm}, very thick,square/.style={regular polygon,regular polygon sides=4,minimum size=0.5cm}, main/.style={draw, circle,minimum size=0.5cm}]

    \node at (0,-0.6) (1) {$\A=\ldots 01010101\ldots$}; 
    \node[right=5.5cm of 1] (3) {$\B=\ldots 01230123\ldots$};
    
    \node at (1.3,1.1) {$a_t$};
    \node at (2.6,-2.5) {$a_t$};
    \node at (7,1.1) {$b_t$};
    \node at (5.7,-2.5) {$b_t$};
	
    \node[ draw, label=above:$\T$] [above right=-0.5cm and 1.4cm of 1] (2) {
    \scalebox{0.7}{\begin{tikzpicture}[node distance={20mm}, very thick, main/.style = {draw, circle,minimum size=0.7cm}] 
		\node[main] (1) [] {$m_0$}; 
		\node[main] (2) [right of=1] {$m_1$}; 
		
		\draw[->] (1) to [out=225,in=135,looseness=5] node[align=center,midway,left] {$0|0:1$ } (1);		
		\draw[->] (2) to [out=45,in=-45,looseness=5] node[align=center,midway,right] {$2|0:1$} (2);		
		\draw[->] (1) to [out=45,in=135,looseness=1] node[align=center,midway,above] {$1|1:1$}  (2); 
		\draw[->] (2) to [out=-135,in=-45,looseness=1] node[align=center,midway,below] {$3|1:1$}  (1); 
	\end{tikzpicture}}
    }; 
    
    \node[draw, label=above:$\hat{\T}$] [below=1.3cm of 2] (8) {
    \scalebox{0.7}{
	\begin{tikzpicture}[node distance={20mm}, very thick, main/.style = {draw, circle,minimum size=0.7cm}] 
		\node[main] (1) {$m^\prime_0$}; 
		\draw[->] (1) to [out=45,in=-45,looseness=5] node[align=center,midway,right] {$0|0:1$\\$1|1:1$\\$0|2:1$\\$1|3:1$ } (1);	 
	\end{tikzpicture} }
    }; 

    \draw[->] (1.center) to [out=90,in=180,looseness=0.8] (2);
	\draw[->] (2) to [out=0,in=90,looseness=0.8] (3);

    \draw[->] (8) to [out=180,in=-90,looseness=0.8] (1);
	\draw[->] (3) to [out=-90,in=0,looseness=0.8] (8);
    \end{tikzpicture} 
    }
\caption{Transformation between processes $\A$ and $\B$. To transform $\A$ to $\B$, memory is necessary, and the memory state $m_j$ is updated as $m_j\leftarrow m_{(j+x)\mod2}$ for the input bit $x$ while the output is generated as $y=2j+x$. On the other hand, to transform $\B$ to $\A$, the output is obtained as $y=x \mod 2$, and thus no memory is necessary. Inside the boxes, we show the finite state machine presentation of the input-output processes. Each transition emitting output $y$ when receiving input $x$, which occurs with probability $p$, is represented as $y|x:p$.}
\label{Fig:scenario}	
\end{figure}
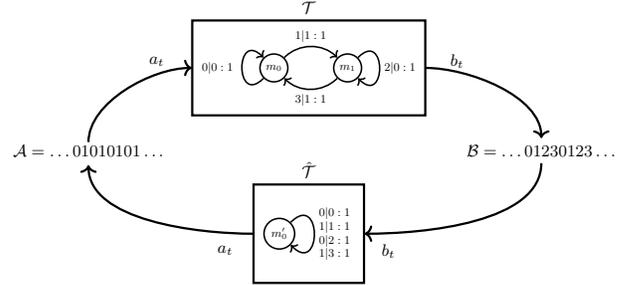

Suppose we observe two streams of data: the first, $\A= \ldots 010101 \ldots$, is an alternating stream of $0$ and $1$s, generated by a binary switch that flips each time-step; the second,  $\B=\ldots 01230123 \ldots$, represents configurations of a revolving object that completes a full cycle every four time-steps. We are asked to decide between two possible causal explanations: (1) $\A$ causes $\B$, such that each $b_t$ of $\B$ at time $t$ is generated by some channel acting on $a_t$ or (2) $\B$ causes $\A$ such that the input and output of this channel are reversed. While we cannot conclusively rule out either causal explanation, the two causal structures are not symmetric. Let $a_t$ and $b_t$ be the respective outputs of $\A$ and $\B$ at time $t$. A simple channel channel enacting $a_t = b_t \mod 2$ transforms $\B$ to $\A$. This channel is memoryless and time-independent. Any agent implementing it does not need to adapt its action to $a_t$ or $b_t$ at times prior to $t$. In contrast, transforming from $\A$ to $\B$ is more complex. An agent that sees $a_t = 0$ cannot decide whether $b_t = 0$ or $b_t = 2$ without at least $1$ bit of information about the past (see Fig.\@ \ref{Fig:scenario}).

In complexity science, each piece of information a machine must track is considered to be a necessary cause of future behaviour~\cite{crutchfield_inferring_1989, barnett_computational_2015}. As such, our thought experiment highlights a variant of \emph{causal asymmetry} - causal explanations in one direction appear more natural than the other~\cite{crutchfield_times_2009, ellison_prediction_2009, thompson_causal_2018}. Here, $\A$ can be thought of as a rudimentary discrete-time clock that tracks 1-bit of data about the current time, while $\B$ is a slightly more sophisticated clock that tracks $2$-bits~\cite{woods2022quantum,yang2018quantum}. Causal asymmetry then represents the intuition that it is easier to degrade a clock than to make it more accurate. Adding noise needs only a memoryless noise channel while reversing this requires adaptive operations.

The goal of this article is to (1) formalize the above intuitions using causal transducers -- a mathematical description of autonomous agents that transform one stochastic process to another~\cite{barnett_computational_2015} and (2) determine how quantum-enhanced agents can change what causal direction we consider to be more natural. This involves defining \emph{\pasm} as the difference in the amount of past data an agent must track to transform $\A$ to $\B$ versus from $\B$ to $\A$; and examining how this difference changes when we deploy agents with quantum memory~\cite{elliott_quantum_2022}. We make a surprising observation: quantum processing can reverse causal asymmetry, such that $\A$ causing $\B$ may be more natural when considering only classical agents, while the reverse is the more natural one when quantum agents are allowed.

\textit{Framework.---} We adopt computational mechanics to describe stochastic processes and agents that transform them from one to another~\cite{crutchfield_between_2012,crutchfield_inferring_1989,barnett_computational_2015}. Formally, a stochastic process $\A$ can be described by a bi-infinite sequence of random variables $\overleftrightarrow{X} = \ldots X_{-1}X_0X_{1}\ldots$, where $X_t$ denotes the probability the process emits $x_t \in \mathcal{X}$ at time $t$. Each instance of a process has a particular past $\overleftarrow{x} = \ldots x_{-1}x_0$, and a conditional future governed by $\overrightarrow{X}|_{\overleftarrow{x}}$. Here we consider stationary stochastic processes, such that $\overleftrightarrow{X}$ is time-translation invariant.

Autonomous agents are finite-state machines that can transform one stochastic process to another~\cite{barnett_computational_2015,elliott_quantum_2022}. At each time-step $t$, an agent accepts input $x_t \in \mathcal{X}$ and emits a corresponding output $y_t \in \mathcal{Y}$. Taking the present as $t = 0$, the agent then experiences a history $\overleftarrow{z}={(\overleftarrow x,\overleftarrow y)}$. The operational behaviour of the agent is defined by an \emph{input-output process}, a family of random variables ${\overrightarrow{Y}|_{\overrightarrow{x},\overleftarrow{z}}}$, which governs the probability the agent will emit $\overrightarrow{y}= y_1y_2 \ldots$ when given future inputs $\overrightarrow{x}= x_1x_2 \ldots$. We say that an agent transforms $\A$ to $\B$ when, given inputs sampled from $\A$, its corresponding outputs $\overleftrightarrow{Y}$ align with $\B$.

When $\overrightarrow{Y}|{\overrightarrow{x},\overleftarrow{z}} = \overrightarrow{Y}|{\overrightarrow{x}}$ is history independent, the agent is \emph{non-adaptive}. This represents the simplest input-output process. Agents implementing such transforms can be \emph{memoryless}. They just need to apply the same transformation on $x_t$ at each time. The set of possible transforms, however, is extremely limited. There is no capability to generate temporal structure, such as converting a sequence of $\A = \ldots0101\ldots$ to $\B = \ldots01230123\ldots$. To do this would require tracking some information about $\overleftarrow{z}$. That is, the agent must possess an internal memory $\mathcal{M}$ that is configured in some state $s_{\overleftarrow{z}} = k(\overleftarrow{z})$, such that there is a systematic method to sample from $\overrightarrow{Y}|_{\overrightarrow{x},\overleftarrow{z}}$ for each possible $\overrightarrow{x}$ and $\overleftarrow{z}$. Here, we assume that all our agents are causal, such that $\mathcal{M}$ does not contain any oracular information - information about the future that is not already contained in the past~\cite{crutchfield2010synchronization,Cabello.PhysRevA.94.052127,thompson_causal_2018}.

To quantify the memory cost of the transformation, we make use of \etrs{}, the provably optimal classical agents for executing a given input-output process~\cite{barnett_computational_2015}. An \etr{} operates on the rationale that an agent needs not distinguish two histories $\overleftarrow{z}$ and $\overleftarrow{z}^\prime$ if its future decisions never depend on this information, i.e. when $\overrightarrow{Y}|_{\overrightarrow{x},\overleftarrow{z}} = \overrightarrow{Y}|_{\overrightarrow{x},\overleftarrow{z}'}$ for all $\overrightarrow{x}$; which we abbreviate to the equivalence relation  $\overleftarrow{z} \er \overleftarrow{z}^\prime$. Thus, \etr{} features an encoding function $k$ such that $k(\overleftarrow{z}) =  k(\overleftarrow{z}')$ if and only if  $\overleftarrow{z} \er \overleftarrow{z}^\prime$. The resulting memory states $\S=\{s_i\}$ are then in $1$-$1$ correspondence with the equivalence classes induced by $\er$. The dynamics of the input-output process can be realized by stochastic transitions between causal states, i.e., a set of transition matrices $\mathcal{T}\equiv \{T^{(y|x)}\}$. Their elements $T^{(y|x)} _{ij}$ then represent the probability an \etr{} in state $s_i$ will emit the symbol $y$ upon receiving input $x$, and transition to state $s_j$. 


The memory states $\{s_i\}$ are referred to as the causal states, owing to their interpretation as the minimal set of belief states an agent must hold to be able to execute $\overrightarrow{Y}|_{\overrightarrow{x},\overleftarrow{z}}$ to perfect statistical fidelity. For each input stochastic process $\A$, the resulting agent memory will be driven to some stationary distribution, $\{\pi_i\}$, where $\pi_i$ denotes the probability of being at causal state $s_i$ when driven by input process $\A$.  
The resulting memory cost $C_\A = -\sum \pi_i \log_2 \pi_i$ is then known as the statistical complexity of the process on input $\A$ and is considered a fundamental measure of structure for $\overrightarrow{Y}|_{\overrightarrow{x},\overleftarrow{z}}$ that characterizes how much past information any agent executing $\overrightarrow{Y}|_{\overrightarrow{x},\overleftarrow{z}}$ must hold in memory. By minimizing the statistical complexity over all input-output processes that transform $\A$ to $\B$, we obtain the minimal memory cost of the transformation, $C_{\A\rightarrow \B}$. 

Note also that this quantity can always be bounded above by $C_{\bf{0}\rightarrow \B}$, the memory cost of transforming a sequence of $\mathbf{0} = \ldots 0000 \ldots$ (or any other i.i.d sequence) to $\B$. This is because we can always choose to first erase $\A$ using a memoryless channel at no cost. The resulting machine is known as the optimal predictive (or causal) model for $\B$~\cite{crutchfield_inferring_1989,crutchfield_between_2012,thompson_causal_2018}. The states of such a machine are referred to as the causal states of $\B$. $C_{\bf{0}\rightarrow \B}$ is referred to as the statistical complexity of $\B$, representing the minimal information we need to store about its past to generate statistically correct future predictions~\cite{crutchfield_inferring_1989}. When  $C_{A\rightarrow \B}$ saturates this bound, it suggests that tracking $\A$ has no benefit for making predictions on $\B$.

Given two processes $\A$ and $\B$, let $C_{\A\rightarrow \B}$ denote the minimal past data needed to transform $\A$ to $\B$, and $C_{\B\rightarrow \A}$ as the equivalent when transforming from $\B$ to $\A$. Then, we define \emph{process causal asymmetry} by the difference
\begin{align}
    \Delta_C(\A,\B) = C_{\A\rightarrow \B}-C_{\B\rightarrow \A} \,.
\end{align}

\noindent A positive $\Delta_C(\A,\B)$ indicates that more memory is required to transform $\A$ to $\B$ than from $\B$ to $\A$. Our opening example illustrates this, with $C_{\A\rightarrow \B} = 1$ and $C_{\B \rightarrow \A} = 0$, such that $\Delta_C(\A,\B) = 1$.

 So far, we assumed that $\mathcal{M}$ stores only classical states. However, quantum agents encoding each causal state $s_i$ to corresponding quantum state $\ket{s_i}$ exhibit a memory advantage~\cite{thompson_using_2017,elliott_quantum_2022}.
The quantum causal states can always be taken to be pure states \cite{elliott_quantum_2022}, and the quantum complexity is defined as the von Neumann entropy $Q_\A = S(\rho_\A) = -\tr \left(\rho_\A \log \rho_\A \right)$ of the average state of the quantum memory, $\rho_\A=\sum_i \pi_{\A,i} \ketbra{s_i}{s_i}$, where $\ketbra{s_i}{s_i}$ is the quantum state representing the classical causal state $s_i$. A quantum model is at least as efficient as the best classical model, i.e., $Q_\A\leq C_\A$ \cite{thompson_using_2017,elliott_quantum_2022}.
By minimizing the quantum complexity over all input-output processes
that transform $\A$ to $\B$, we obtain $Q_{\A\rightarrow \B}$, the minimal memory needed by quantum agents to transform $\A$ to $\B$. For certain $\A$ and $\B$, we have that $Q_{\A\rightarrow \B} <  C_{\A\rightarrow \B}$. 
In analogy to the classical case, we define process causal asymmetry for quantum agents as
$$\Delta_Q(\A,\B) =  Q_{\A\rightarrow \B}-Q_{\B\rightarrow \A}.$$

\emph{Main Results.---} We now state our main results. The first shows that the direction of causal asymmetry can be reversed by quantum agents. 
\begin{result}[Inconsistent Causal Asymmetry]
There exist processes $\A$ and $\B$ with process causal asymmetries simultaneously obeying $\Delta_C(\A,\B)>0$ and $\Delta_Q(\A,\B)<0$, which implies that the classical memory resources necessary to map $\A$ to $\B$ are higher than that of mapping $\B$ to $\A$, while quantum mechanically it is the other way around. Causal asymmetry can thus reverse depending on whether quantum agents are allowed or not.   
\end{result}

Our second result shows that not only an inconsistent causal asymmetry exists, but the magnitude of this inconsistency can be arbitrarily large.
\begin{result} [Unbounded Inconsistent Causal Asymmetry]
    There exist processes $\A$ and $\B_n$, for $n \in \mathbb{N}$ such that the classical process asymmetry $\Delta_C(\A,\B)$ diverges with increasing $n$, while the quantum one, $\Delta_Q(\A,\B)$, remains bounded but with the opposite sign.
\end{result}

Our results thus illustrate - for the first time to our knowledge - that quantum and classical theories can result in different conclusions about the most natural order of causal transformations. 
The possibility of storing information on non-orthogonal quantum states, not only leads to a quantitative reduction of memory cost \cite{thompson_using_2017,elliott_quantum_2022}, but to a qualitatively different behaviour in the direction of causation.

\emph{Explicit Demonstration --} Let us consider two processes that consist of a ground state to which the probability of transition is always $\nicefrac{1}{2}$ from any of the excited states. The first process has only one excited state and is a coarse-graining of the second one, which has multiple excited states.

\begin{figure}[!t]
    \centering
    \scalebox{0.75}{
	\begin{tikzpicture}[envel/.style={shape=ellipse, draw, inner sep=0.35cm, minimum height=2.15cm},node distance={15mm}, very thick, square/.style={regular polygon,regular polygon sides=4,minimum size=0.5cm}, main/.style = {draw, circle,minimum size=0.7cm},darkstyle/.style={circle,draw,fill=gray!20,minimum size=0.7cm},bluestyle/.style={circle,draw,fill=blue!20,minimum size=0.7cm},greenstyle/.style={circle,draw,fill=green!20,minimum size=0.7cm}] 

        \node[darkstyle] (a)  {$r_0$} ; 
		\node[bluestyle] (b) [above= 2.5cm of a] {$r_1$}; 
        \node (c) [scale=1.3, above left=0.5cm and 0.5cm of b] {\framebox{$\A$}};
        
		\draw[blue, ->] (a) to [out=45,in=-45,looseness=0.7] node[align=center,midway,right] {$1:\nicefrac{1}{2}$\\$2:\nicefrac{1}{2}$}  (b); 
		\draw[red, ->] (b) to [out=-135,in=135,looseness=0.7] node[align=center,midway,left] {$0:\nicefrac{1}{2}$}  (a); 
		\draw[->] (b) to [out=135,in=45,looseness=5] node[align=center,midway,above] {$1:\nicefrac{1}{2}$} (b);
	   
        \node[] (5) [right= 3cm of b] {};
        \node[] (6) [right = 4.5cm of 5] {};
        \node[fit=(5)(6), fill=yellow!20, envel] {};
        
		\node[bluestyle] (1) [right = 0cm of 5] {$s_j$}; 
        \node (10) [scale=1.3,above left=0.5cm and 1.cm of 1] {\framebox{$\B_n$}};
		\node[greenstyle] (4) [below right=1cm and 1.275cm of 1] {$s_2$}; 
        \node[darkstyle] (3) [below=1cm of 4] {$s_0$}; 
        \node[bluestyle] (2) [right=3cm of 1] {$s_{k}$};

        \node (11) [above right=0.75cm and 2.5cm of 1] {$j,k=1,3,\ldots,n-1$};
        \node[] (7) [right= 1.5cm of 2] {};
        \node[] (8) [left=1.5cm of 1] {};
           
        \path (1) -- (2) node [font=\huge, midway, sloped] {$\dots$};
        \path (2) -- (7) node [font=\huge, midway, sloped] {$\dots$};
        \path (1) -- (8) node [font=\huge, midway, sloped] {$\dots$};
            
        \draw[red, ->] (1) to [out=-125,in=165,looseness=0.9] node[align=center,midway,right,pos=0.56] {$0:\frac{1}{2}$} (3);
        \draw[red, ->] (2) to [out=-55,in=15,looseness=0.9] node[align=center,left,pos=0.45] {} (3);

        \draw[->] (3) to [out=180, in=-145,looseness=0.85] node[align=center,midway,below =0.1cm, left ] {} (1);
        \draw[->] (3) to [out=0,in=-35,looseness=0.85] node[align=center,right,pos=0.31] {} (2);
        \draw[blue,->] (3) to [out=75, in=-75,looseness=0.9] node[align=center,midway,below =0.1cm, right,pos=0.8] {$2:\frac{1+p_{02}}{2}$} (4);
           
        \draw[->] (1) to [out=30,in=150,looseness=0.7] node[align=center,midway,above,pos=0.7] {$k:\frac{p_{jk}}{2}$} (2);
        \draw[->] (1) to [out=140,in=80,looseness=5] node[align=center,left,pos=0.1] {$j:\frac{p_{jj}}{2}$} (1);
        \draw[->] (1) to [out=-90, in=180,looseness=0.9] node[align=center,midway,below =0.1cm, pos=0.5] {} (4);

        \draw[->] (2) to [out=210,in=-30,looseness=0.7] node[align=center,midway,above,left] {} (1);
        \draw[->] (2) to [out=100,in=40,looseness=5] node[align=center,midway,above] {} (2);
        \draw[->] (2) to [out=-90, in=0,looseness=0.9] node[align=center,midway,below =0.1cm, left ] {} (4);

        \draw[->] (4) to [out=95,in=0,looseness=0.7] node[align=center,midway,above,left] {} (1);
        \draw[->] (4) to [out=85,in=180,looseness=0.7] node[align=center,midway,above,left] {} (2);
        \draw[->] (4) to [in=145,out=-145,looseness=6] node[align=center,midway,above] {} (4);
        \draw[red, ->] (4) to [out=-110,in=110,looseness=0.9] node[align=center,midway,above,left] {} (3);
    \end{tikzpicture}}

\caption{Processes $\A$ and $\B_n$. Labels of transitions with the same colour share the same pattern.}
	\label{Fig: processes A and B}
\end{figure}
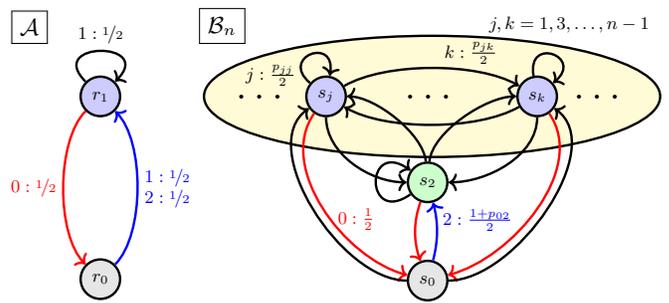

In detail, let $\A$ be a process with alphabet $x\in \X = \{0,1,2\}$, whose statistics can be generated by a machine with two internal states, ground state $r_0$ and excited state $r_1$ (see Fig.~\ref{Fig: processes A and B}). An output of $x=0$ is followed by a transition to state $r_0$, while outputs $x=1$ or $x=2$ lead to a trasition to state $r_1$, with probability $\nicefrac{1}{2}$.

We also consider a family of stochastic processes, $\B_n$, labeled by an index $n\in \mathbb{N}$ with $n\geq3$.  Each can be modeled by a machine with $n$ causal states, $s_0\,,\ldots,s_{n-1}$, and $n$ outputs, $y\in \Y =\{0,\ldots,n-1\}$ with the following properties: (i) an emission of symbol $y_j$ leads to a transition to the state of the same index $s_{y_j}$, (ii) any excited state $s_{j \neq 0}$ either transitions to the ground state $s_0$ with probability $\nicefrac{1}{2}$ or transitions to another excited state $s_{k \neq 0}$ with probability $\nicefrac{p_{jk}}{2}$, (iii) $s_0$ transitions to one of the excited states $s_{j \neq 0,2}$ with probability $\nicefrac{p_{0j}}{2}$, except for $j=2$ which occurs with probability $\nicefrac{1+p_{02}}{2}$. Moreover, the probabilities $p_{jk}$, satisfying $\sum_{k =1}^{n-1} p_{jk} = 1$ for all $j$, are all assumed to be close to the value $\nicefrac{1}{n-1}$ but different from each other. These processes are shown in Fig.\@ \ref{Fig: processes A and B}.

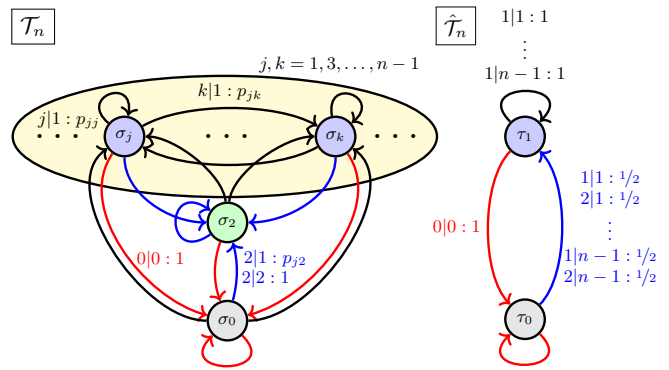
\begin{figure}[!b]
    \centering
    \scalebox{0.75}{
	\begin{tikzpicture}[envel/.style={shape=ellipse, draw, inner sep=0.35cm, minimum height=2.15cm},node distance={15mm}, very thick, square/.style={regular polygon,regular polygon sides=4,minimum size=0.5cm}, main/.style = {draw, circle,minimum size=0.7cm},darkstyle/.style={circle,draw,fill=gray!20,minimum size=0.7cm},bluestyle/.style={circle,draw,fill=blue!20,minimum size=0.7cm},greenstyle/.style={circle,draw,fill=green!20,minimum size=0.7cm}] 

        \node[darkstyle] (a)  {$\tau_0$} ; 
		\node[bluestyle] (b) [above= 2.5cm of a] {$\tau_1$}; 
        \node (c) [scale=1.2, above left=1.2cm and 0.45cm of b] {\framebox{$\hat{\T}_n$}};
        
		\draw[blue, ->] (a) to [out=45,in=-45,looseness=0.7] node[align=center,pos=0.5,right=-0.1cm] {$1|1:\nicefrac{1}{2}$\\$2|1:\nicefrac{1}{2}$\\ \vdots \\$1|n-1:\nicefrac{1}{2}$\\$2|n-1:\nicefrac{1}{2}$}  (b); 
		\draw[red, ->] (b) to [out=-135,in=135,looseness=0.7] node[align=center,midway,left] {$0|0:1$}  (a); 
		\draw[->] (b) to [out=135,in=45,looseness=5] node[align=center,midway,above] {$1|1:1$\\ \vdots \\$1|n-1:1$} (b);
        \draw[red, ->] (a) to [out=-45,in=-135,looseness=5] node[align=center,midway,below] {} (a);
	   
        \node[] (5) [left= 7.cm of b] {};
        \node[] (6) [right = 4.1cm of 5] {};
        \node[fit=(5)(6), fill=yellow!20, envel] {};
        
		\node[bluestyle] (1) [right = -0.1cm of 5] {$\sigma_j$}; 
        \node (10) [scale=1.3, above left=1.2cm and 0.8cm of 1] {\framebox{$\T_n$}};
		\node[greenstyle] (4) [below right=1cm and 1.3cm of 1] {$\sigma_2$}; 
        \node[darkstyle] (3) [below=1cm of 4] {$\sigma_0$}; 
        \node[bluestyle] (2) [right=3cm of 1] {$\sigma_{k}$};

        \node (11) [above right=0.75cm and 2cm of 1] {$j,k=1,3,\ldots,n-1$};
        \node[] (7) [right= 1.5cm of 2] {};
        \node[] (8) [left=1.5cm of 1] {};
           
        \path (1) -- (2) node [font=\huge, midway, sloped] {$\dots$};
        \path (2) -- (7) node [font=\huge, midway, sloped] {$\dots$};
        \path (1) -- (8) node [font=\huge, midway, sloped] {$\dots$};
            
        \draw[red, ->] (1) to [out=-125,in=165,looseness=0.9] node[align=center,midway,right,pos=0.56] {$0|0:1$} (3);
        \draw[red, ->] (2) to [out=-55,in=15,looseness=0.9] node[align=center,left,pos=0.45] {} (3);

        \draw[->] (3) to [out=180, in=-145,looseness=0.85] node[align=center,midway,below =0.1cm, left ] {} (1);
        \draw[->] (3) to [out=0,in=-35,looseness=0.85] node[align=center,right,pos=0.31] {} (2);
        \draw[blue,->] (3) to [out=75, in=-75,looseness=0.9] node[align=center,midway,below =0.1cm, right,pos=0.65] {$2|1:p_{j2}$\\$\hspace{-0.34cm}2|2:1$} (4);
        \draw[red, ->] (3) to [out=-45,in=-135,looseness=5] node[align=center,midway,below] {} (3);
           
        \draw[->] (1) to [out=30,in=150,looseness=0.7] node[align=center,midway,above,pos=0.5] {$k|1:p_{jk}$} (2);
        \draw[->] (1) to [out=140,in=80,looseness=5] node[align=center,left=-0.05,pos=0.05] {$j|1:p_{jj}$} (1);
        \draw[blue, ->] (1) to [out=-90, in=180,looseness=0.9] node[align=center,midway,below =0.1cm, pos=0.5] {} (4);

        \draw[->] (2) to [out=210,in=-30,looseness=0.7] node[align=center,midway,above,left] {} (1);
        \draw[->] (2) to [out=100,in=40,looseness=5] node[align=center,midway,above] {} (2);
        \draw[blue, ->] (2) to [out=-90, in=0,looseness=0.9] node[align=center,midway,below =0.1cm, left ] {} (4);

        \draw[->] (4) to [out=95,in=0,looseness=0.7] node[align=center,midway,above,left] {} (1);
        \draw[->] (4) to [out=85,in=180,looseness=0.7] node[align=center,midway,above,left] {} (2);
        \draw[blue, ->] (4) to [in=145,out=-145,looseness=6] node[align=center,midway,above] {} (4);
        \draw[red, ->] (4) to [out=-110,in=110,looseness=0.9] node[align=center,midway,above,left] {} (3);
    \end{tikzpicture}}

\caption{The \etrs{} $\T$ and $\hat{\T}_n$. Labels of transitions with the same colour share the same pattern.}
	\label{Fig: etrs}
\end{figure}

We now turn to \etrs{} that transform $\A$ to $\B_n$. Briefly speaking, such an agent has one state for each different state of the output $\B_n$ and modifies the probabilities of the outputs of $\A$ accordingly in order to reproduce the probabilities of $\B_n$. These constraints alone are not sufficient to single out a unique construction, and thus, there exists a family of \etrs{} that take $\A$ to $\B_n$ with minimal complexity. Nevertheless, as they all have the same classical complexity (see Appendix A), we choose a representative \etr{} $\T_n$ with $n$ states, $\Sigma=\{\sigma_i\}_{i=0\ldots n-1}$, a 3-symbol input alphabet $\X=\{0,1,2\}$, and an $n$-symbol output alphabet $\Y=\{0,\ldots,n-1\}$ (see Fig.\@ \ref{Fig: etrs}). 
Whenever an emission of a certain symbol is made, a transition to the internal state of the same label occurs. Specifically, if a symbol $x\neq 1$ is input, the same symbol $y=x$ is output, and a transition to state $\sigma_x$ is made, respectively. For input 1, however, output $y=j$ is emitted with probability $p_{ij}$ if the \etrs{} are at state $\sigma_i$. 

Similarly, the \etrs{} that transform $\B_n$ to $\A$ keep intact the emissions of symbol $0$ but have to `erase' all the probabilities associated with the remaining $n-1$ outputs. We denote this with $\hat{\T}_n$, and a graphic representation is shown in Fig.\@ \ref{Fig: etrs}. Once again, more than one channel yielding the same transformation exists, but they all have the same classical statistical complexity (see Appendix A).

We now show that the classical complexities $C_{\A\rightarrow \B_n}$ of the \etrs{} $\T_n$ that transform from $\A$ to $\B_n$ are increasing with $n$, while the quantum complexities $Q_{\A\rightarrow \B_n}$ get arbitrarily close to 0 when the probabilities $p_{ij}$ get closer. In the other direction, however, both classical, $C_{\B_n\rightarrow \A}$, and quantum, $Q_{\B_n\rightarrow \A}$, complexities assume a finite value between 0 and 1. 

We start with the classical complexity of the \etr{} $\hat{\T}_n$ from $\A$ to $\B_n$. We first obtain the stationary distribution of the process $\A$ itself and then the long-term probabilities of symbol $x$ being emitted.
We explicitly find the values $\left\{\Pr(0),\Pr(1),\Pr(2)\right\}=\left\{\nicefrac{1}{3},\nicefrac{1}{2},\nicefrac{1}{6}\right\}$. By also noting that every output of the \etrs{} leads to a transition to the state with the same label, the stationary distribution of the \etrs{} can be evaluated (see Appendix B).
For $n\geq3$ and under the assumption $p_{i,j}\approx \nicefrac{1}{n-1} ~\forall i,j,$ we find the stationary distribution $\varphi_0 = \Pr(0), \varphi_2 = \nicefrac{\Pr(1)}{n-1} +\Pr(2)$, and $\varphi_j = \nicefrac{\Pr(1)}{n-1}$, for $j=1,3,\ldots,n-1$. The complexity is obtained as $C_{\A\rightarrow \B_n}=-\sum_{i} \varphi_i \log \varphi_i$; in the limit of large $n$ it grows logarithmically with $n$, i.e., $C_{\A\rightarrow \B_n}\sim \log n$, and thus it is unbounded.

On the other hand, the \etrs{} that map $\B_n$ to $\A$ can be shown to have a classical complexity independent of $n$. The stationary distribution is $\hat{\varphi}_0=\Pr(0) = 1/3 \,, \hat{\varphi}_1 = 1-\Pr(0) = 2/3$, where now $\Pr(0)$ is the probability of process $\B_n$ emitting a 0. However, since $\T_n$ preserved the emissions with 0, $\Pr(0)$ for $\B_n$ is the same as that of the input $\A$, from which we obtain $C_{\B_n\rightarrow \A}= -\sum_{i=0,1} \hat{\varphi}_i \log \hat{\varphi}_i \approx 0.918$ independent of $n$. It follows that the classical process causal asymmetry is 
$$
\Delta_C(\A, \B_n) \sim \log n,
$$
which grows as the logarithm of $n$. In other words, transforming process $\A$ to $\B_n$ is increasingly costly in terms of memory when a classical memory is employed. 

We now turn to the derivation of the quantum complexity, $Q_{\A\rightarrow \B_n}$. We explicitly construct the following quantum models where the classical causal states are encoded as $\ket{s_i}=\otimes_x\ket{s_i^x}$ with $\ket{s_i^x}=\sum_{k}\sum_{y} \sqrt{T_{ik}^{(y|x)}} \ket{y}\ket{k}$~\cite{elliott_quantum_2022,thompson_causal_2018}.
The \etrs{} $\T_n$ are such that for inputs $x=0,2$ we have that $\ket{s_i^x}=\ket{x}\ket{x}$ independent of the state index $i$. Also, one can easily check that $\ket{s_i^1}=\sum_{j\neq 0} \sqrt{p_{ij}} \ket{j}\ket{j}$. As a result, the overlap between causal states is given by $\braket{s_i}{s_j} = \sum_{k\neq 1} \sqrt{p_{ik}\, p_{jk}}$. The quantum complexity is the von Neumann entropy of the state $\rho = \sum_i \varphi_i \ketbra{s_i}{s_i}$ where $\varphi_i$ denotes the $i-$th element of the stationary distribution of the \etr{} $\T_n$ when driven by the input $\A$. From the fact that state $\rho$ and the Gram matrix of overlaps, with elements $G_{ij}=\sqrt{\varphi_i \varphi_j} \braket{s_i}{s_j}$, have the same non-zero eigenvalues \cite{jozsa_distinguishability_2000}, thereby their von Neumann entropies coincide. When all the probabilities $p_{ij}$ get arbitrarily close to each other, we have that $G_{ij}\approx \sqrt{\varphi_i \varphi_j}$.
In this case, the Gram matrix $G\approx v v^\top $, with $v^\top = (\sqrt{\varphi_1}\,, \ldots \,, \sqrt{\varphi_n} )$  has approximately one eigenvalue equal to 1 and all others equal to 0, leading to a quantum complexity of 0. In Appendix C we give a precise argument and provide an upper bound for a small perturbation of the probabilities around the value $\nicefrac{1}{n-1}$ so that $p_{ij}=\frac{1+\delta_{ij}}{n-1}$, with $\abs{\delta_{ij}}\leq \delta$ for some small $\delta>0$, and we show that the optimal quantum complexity is bounded as
\begin{align}
    Q_{\A\rightarrow \B_n}^{\text{upper}}(\delta) \geq Q_{\A\rightarrow \B_n}\geq 0\,, \label{eq:A to B}
\end{align}
where $Q_{\A\rightarrow \B_n}^{\text{upper}}(\delta)$ can be made arbitrarily small by an appropriate choice of $\delta$. 

\begin{figure}[!t]
	\centering	\includegraphics[width=0.95\linewidth]{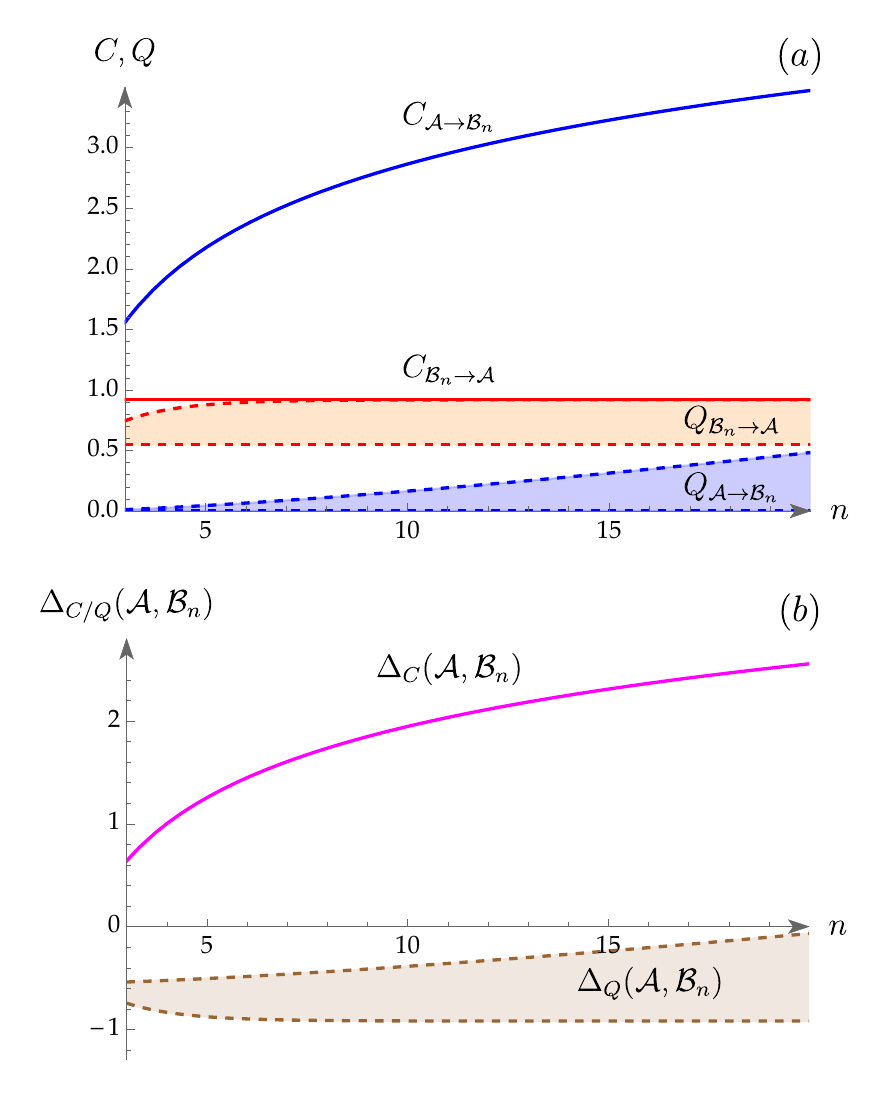}
	\caption{(a) Classical complexities $C_{\A\rightarrow \B_n}$ (blue, solid line) and $C_{\B_n \rightarrow \A}$ (red, solid line), as well as the quantum complexities, $Q_{\A\rightarrow \B_n}$ (light blue shaded area) and $Q_{\B_n \rightarrow \A}$ (orange shaded area). (b) Classical and quantum process causal asymmetries between $\A$ and $\B_n$. In magenta we show $\Delta_{C}(\A,\B_n)$, while $\Delta_{Q}(\A,\B_n)$ lies in the brown-shaded area.}
	\label{Fig:first few cases approximation}
\end{figure}

Turning to the quantum complexity of the \etrs{} from $B_n$ to $\A$, the situation is more intricate. We consider all possible classical channels and show that the complexity of all possible quantum models is bounded (see Appendix A). This follows from a maximal fidelity constraint that any model has to obey to correctly reproduce the channel \cite{elliott_quantum_2022}. Specifically, we can show that the optimal quantum complexity is bounded from below and above according to the inequality
\begin{align}
    Q_{\B_n\rightarrow \A}^{\text{upper}} \geq Q_{\B_n\rightarrow \A}\geq 0.55\,, \label{eq:B to A}
\end{align}
where $Q_{\B_n\rightarrow \A}^{\text{upper}}$ depends on the value of $n$; it takes the value 0.682 for $n=3$ and approaches the classical complexity of 0.918 with increasing values of $n$. 

Using Eqs.\@ \eqref{eq:A to B} and \eqref{eq:B to A}, we obtain the following bounds on the quantum process causal asymmetry, $\Delta_Q(\A,\B_n)$,
\begin{align}
    Q^{\text{upper}}_{\A\rightarrow \B_n} (\delta)-0.55  \geq  \Delta_Q(\A,\B_n) \geq - Q_{\B_n\rightarrow \A}^{\text{upper}}.
\end{align}
On the left hand side and for any $n$, there exists a range of $\delta$ that make this difference negative (see Appendix A). This implies that mapping processes $\B_n$ to $\A$ has a higher memory cost than that of mapping $\A$ to $\B_n$, when a quantum memory is employed. 

In Fig. \ref{Fig:first few cases approximation}, we plot the difference of classical and quantum process causal asymmetries, 	$\Delta_C(\A,\B_n)$ and $
\Delta_Q(\A,\B_n)$, for $n=3,\ldots,20$ and $\delta=10^{-2}$. We see that the classical one grows logarithmically with $n$ while the quantum approaches a finite negative value.

\textit{Discussion.---}
Transforming one stochastic process into another generally requires memory. Here, we introduced \emph{process causal asymmetry} to investigate the difference in memory costs required for transformations between a pair of stochastic processes $\A$ and $\B$. We demonstrated that agents with quantum memory exhibit dramatically different behaviour from those with classical memory. In particular, we provided the first explicit example where the direction of causation is flipped between classical and quantum models, and the gap grows unboundedly with system size.

There are several natural directions for further research. The first involves linking our work to that of causal asymmetry in the context of causal vs.\@ retrocausal models -- where memory costs of modeling stochastic processes in forward-time vs. reverse-time can differ~\cite{crutchfield_times_2009}. Here, quantum models could remove this asymmetry, but a full reversal remains unknown~\cite{thompson_causal_2018}. Meanwhile, our quantum agents transform classical data to classical data - while quantum agents (combs) that transform quantum processes also exist~\cite{paparo2014quantum, breuer2016colloquium, chiribella2009theoretical}. It would be exciting to unify such settings, considering causal asymmetry when both time and input-outputs are reversed in contexts of both classical and quantum outputs.

Conceptually, process causal asymmetry also hints at some underlying resource theory on temporal correlations. For example, a sequence of i.i.d. random variables would take no memory to generate, while transforming this to a highly non-Markovian process requires adaptive agents with increasing memory resources~\cite{yang2020ultimate, vieira2022temporal}. This opens an interesting research direction towards constructing a formal hierarchy of non-adaptive agents and relating it to existing ideas from resource theories of non-Markovianity~\cite{berk2021resource, anand2019quantifying}. Indeed, the power of memory and adaptive operations has already been recognized in quantum state preparation~\cite{v001a005, bravyi2022adaptive}, gate synthesis~\cite{Jones13, Gidney2018halvingcostof}, exhibiting contextuality~\cite{Cabello.PhysRevA.94.052127,kleinmann2011memory,budroni2019contextuality,cabello2018optimal}, and generating data-strings~\cite{vieira2022temporal} as well as being a key differentiator between quantum error mitigation and error correction~\cite{takagi2022fundamental}. Therefore, this direction could provide valuable insights on when quantum agents have greatest advantage and where they are most useful.


\textit{Acknowledgements.---}
We thank Paul Riechers, Jayne Thompson, Thomas Eliott, Andrew Garner, Felix Binder and Alec Boyd for interesting discussions. This work is supported by the National Research Foundation of Singapore, and the Agency for Science, Technology and Research (A*STAR) under its QEP2.0 programme (NRF2021-QEP2-02-P06), the Singapore Ministry of Education Tier1 Grant RG77/22, the Singapore Ministry of Education Tier 2 Grant MOET2EP50221-0005, FQxI under grant nos.~FQXi-RFP-IPW-1903 (``Are quantum agents more energetically efficient at making predictions?") and FQXi-RFP-1809 (``The role of quantum effects in simplifying adaptive agents") from the Foundational Questions Institute and Fetzer Franklin Fund (a donor-advised fund of Silicon Valley Community Foundation).
H.K. and S.K. are supported by KIAS Individual Grant Nos.~CG085301 (H.K.) and CG086201 (S.K.) at the Korea Institute for Advanced Study.

\bibliography{IO_causal_asymmetryBib}
\bibliographystyle{apsrev4-2}	

\newpage~\newpage
	
\appendix

\section{Appendix A: Derivation of minimal \etrs{} $\T_{n}$ and $\hat{\T}_{n}$} 
We derive all minimal \etrs{} from process $\A$ to process $\B_n$ and vice versa. We go through all the details of the calculation for the \etrs{} $\hat{\T}_n$ from $\B_n$ to $\A$ and only describe the general characteristics the other way around. Although we show the derivation explicitly in the case with $n=3$, the proof remains valid for any $n$ and comment on the general case in the end. 

Since each output of $\A$ and $\B_3$ synchronises to a unique state \cite{travers_exact_2011}, this needs to be reflected in the structure of the \etr{}. As a result, regardless of input, outputs of 1 or 2 need to synchronize to state $r_1$ say of the \etr{}, while conditional outputs of 0, need to synchronize to the other state, $r_0$. In practice, this also suffices to show that no memoryless \etr{} that maps $\B_3$ to $\A$ exists, but we also show it explicitly. 
Assume that such an \etr{} exists. In this case, transition matrices are just a collection of probabilities since there is only one state and have the general form
\begin{align}
    \hat{T}:\{&p_{00},\, p_{10},\, p_{20}=1-p_{00}-p_{10}\, \notag \\
    &p_{01},\, p_{11},\, p_{21}=1-p_{01}-p_{11}\, \notag \\
    &p_{02},\, p_{12},\, p_{22}=1-p_{02}-p_{12}\} \,,
\end{align}
where the symbols $p_{ij}$ have subscripts that are shorthand for $i|j$ denoting the probability of outputting symbol $i$ while receiving symbol $j$.

The transition matrices of a process $\A$ are
\begin{align}
    A^{(x)} = \left\{ \begin{pmatrix}
        0 & 0 \\
        \nicefrac{1}{2} & 0
    \end{pmatrix},
    \begin{pmatrix}
        0 & \nicefrac{1}{2} \\
        0 & \nicefrac{1}{2}
    \end{pmatrix},
     \begin{pmatrix}
        0 & \nicefrac{1}{2} \\
        0 & 0
    \end{pmatrix}
    \right\}\,,
\end{align}
with $x=0,1,2$ and similarly for $\B_3$:
\begin{align}
    B_3^{(x)} = \left\{ \begin{pmatrix}
        0 & 0 & 0 \\
        \nicefrac{1}{2} & 0 & 0 \\
        \nicefrac{1}{2} & 0 & 0 \\
    \end{pmatrix},
    \begin{pmatrix}
        0 & \nicefrac{p_0}{2} & 0 \\
        0 & \nicefrac{p_1}{2} & 0 \\
        0 & \nicefrac{p_2}{2} & 0 \\
    \end{pmatrix},
    \begin{pmatrix}
        0 & 0 & 1-\nicefrac{p_0}{2} \\
        0 & 0 & \nicefrac{(1-p_1)}{2} \\
        0 & 0 & \nicefrac{(1-p_2)}{2} \\
    \end{pmatrix}
    \right\}\,,
\end{align}
where $p_i=p_{i1}$.

Taking the process $\B_3$ with transition matrices $B_3^{(x)}$ as input to the \etr{} $\hat{\T}_3$ with transition matrices $\hat{T}^{(y|x)}$ one can construct the transition matrices $J^{(x,y)}$ of the joint process over joint inputs and outputs as
\begin{align}
    J^{(x,y)}=\hat{T}^{(y|x)} \otimes B_3^{(x)} \,.
\end{align}
This follows by noting that element-wise this is equivalent to the statement about conditional probabilities
\begin{align}
   J^{(x,y)}_{(i,j)\rightarrow (k,l)}
   &=\Pr(Y_{t}=y, X_t =x, S_{t+1}=s_k, \notag\\
   &\,\,\,\,\,\,\,\,\,\,\,\,\,\,\,\,\,\,\,\,\Sigma_{t+1}=\sigma_l |  , S_{t}=s_i, \Sigma_{t}=\sigma_j)  \notag \\
   &=\Pr(Y_{t}=y, S_{t+1}=s_k |  X_t =x, S_{t}=s_i)\cdot \notag \\
   &\,\,\,\,\,\,\,\,\,\,\,\,\,\,\, \Pr( \Sigma_{t+1}=\sigma_l, X_t =x |\Sigma_{t}= \sigma_j)
\end{align}
From the joint process, one can trace over inputs, $x$, to obtain a presentation of the process over outputs alone:
\begin{align}
  \tilde{A}^{(y)} =  \sum_x J^{(x,y)} \,.
\end{align}
We put a tilde to the output process obtained this way to signify the fact that in general this will not be a minimal presentation and some of the states of the output may be \emph{transient}, that is, have long term probability of occurrence equal to zero. Moreover, some states may also be equivalent to others, that is, belong to the same equivalence class.  In order to obtain the minimal presentation, the \ema{}, of the output process, one needs to remove transients and merge equivalent states. Both processes $\A$ and $\B_3$ are given in their minimal presentation and thus upon removing transient states and merging equivalent states, the resulting matrices of the output should be the same as those of $\A$.  Explicitly, for output $y=0$ we have
\begin{align}
    \tilde{A}^{(0)} &= \sum_x \hat{T}^{(0|x)} \otimes B_3^{(x)} = \sum_x p_{0x} B_3^{(x)} \notag \\
    &= p_{00} B^{(0)}+p_{01} B^{(1)}+p_{02} B^{(2)} \notag \\
    &=\frac{1}{2}\begin{pmatrix}
        0 & p_{01} p_0 & p_{02}(2-p_0) \notag \\
        p_{00} & p_{01} p_1 & p_{02}(1-p_1) \notag \\
        p_{00} & p_{01} p_2 & p_{02}(1-p_2) 
        \end{pmatrix}
\end{align}
Examining the structure of $A^{(0)}$, the target output, and the output $\tilde{A}^{(0)}$ of the memoryless \etr{}, we see that the only possibility is that states 2 and 3 of $\tilde{A}$ are in the same equivalence class, call it class 2, corresponding to the second state of the target process $\A$. Moreover, since there are no transitions from class 1 to class 2, we must necessarily have $p_{01}=p_{02}=0$. Similarly, for output $x=2$ we have
\begin{align}
    \tilde{A}^{(2)} 
    &= p_{20} B^{(0)}+p_{21} B^{(1)}+p_{22} B^{(2)} \notag \\
    &=\frac{1}{2}\begin{pmatrix}
        0 & p_{21} p_0 & p_{22}(2-p_0) \notag \\
        p_{20} & p_{21} p_1 & p_{22}(1-p_1) \notag \\
        p_{20} & p_{21} p_2 & p_{22}(1-p_2). 
    \end{pmatrix}
\end{align}
Since at equivalence class 2 there are no self transitions while emitting symbol 2, we must have that $p_{21}=p_{22}=0$, which however implies that no transition from class 2 to class 1 can happen. This shows that the output can not be correctly reproduced with a memoryless \etr{}.
Thus the \etr{} must have at least two states. 

A general Markovian on outputs (more precisely, of \emph{pure feedback Markov order} 1 \cite{barnett_computational_2015}) \etr{} with two states has the following transition matrices
\begin{align}
    \hat{T}:\{&\begin{pmatrix}
        t^{00}_{11} & 0 \\
        t^{00}_{21} & `
    \end{pmatrix}\,,
    \begin{pmatrix}
        0 & t^{10}_{12} \\
        0 & t^{10}_{22}
    \end{pmatrix}\,,
     \begin{pmatrix}
        0 & t^{20}_{12} \\
        0 & t^{20}_{22}
    \end{pmatrix}\,,  \notag \\
    &\begin{pmatrix}
        t^{01}_{11} & 0 \\
        t^{01}_{21} & 0
    \end{pmatrix}\,,
    \begin{pmatrix}
        0 & t^{11}_{12} \\
        0 & t^{11}_{22}
    \end{pmatrix}\,,
     \begin{pmatrix}
        0 & t^{21}_{12} \\
        0 & t^{21}_{22}
    \end{pmatrix}\,, \notag \\
    &\begin{pmatrix}
        t^{02}_{11} & 0 \\
        t^{02}_{21} & 0
    \end{pmatrix}\,,
    \begin{pmatrix}
        0 & t^{12}_{12} \\
        0 & t^{12}_{22}
    \end{pmatrix}\,,
     \begin{pmatrix}
        0 & t^{22}_{12} \\
        0 & t^{22}_{22}
    \end{pmatrix}\} \,,
\end{align}
where in the matrix elements, $t_{ij}^{yx}$, superscripts stand for conditional outputs given inputs, $y|x$, and subscripts $ij$ denote transition from causal state $i$ to causal state $j$.
Conservation of probability and unifilarity imposes the constraints
\begin{align}
    t^{00}_{11}+t^{10}_{12}+t^{20}_{12}&=1 \,, \, \,  t^{00}_{21}+t^{10}_{22}+t^{20}_{22}=1 \,, \notag \\
    t^{01}_{11}+t^{11}_{12}+t^{21}_{12}&=1 \,, \, \, 
    t^{01}_{21}+t^{11}_{22}+t^{21}_{22}=1 \,, \notag \\
    t^{02}_{11}+t^{12}_{12}+t^{22}_{12}&=1 \,, \, \,
    t^{02}_{21}+t^{12}_{22}+t^{22}_{22}=1 \,. \notag \\
\end{align}
We calculate the output of the \etr{} when driven with $\B_3$ by using the composition and marginalization equation,
\begin{align}
    \tilde{A}^{(y)} &= \sum_x \hat{T}^{(y|x)} \otimes B_3^{(x)}\,.
\end{align}
Explicitly we obtain the 6 state output process with transition matrices:
\begin{align}
    \tilde{A}^{(0)} = \frac{1}{2}\begin{pmatrix}
        0 & t^{01}_{11} p_0 & t^{02}_{11}(2-p_0) & 0 & 0 & 0 \\
        t^{00}_{11} & t^{01}_{11} p_1 & t^{02}_{11}(1-p_1) & 0 & 0 & 0 \\
        t^{00}_{11} & t^{01}_{11} p_2 & t^{02}_{11}(1-p_2) & 0 & 0 & 0 \\
        0 & t^{01}_{21} p_0 & t^{02}_{21}(2-p_0) & 0 & 0 & 0 \\
        t^{00}_{21} & t^{01}_{21} p_1 & t^{02}_{21}(1-p_1) & 0 & 0 & 0 \\
        t^{00}_{21} & t^{01}_{21} p_2 & t^{02}_{21}(1-p_2) & 0 & 0 & 0 \\
    \end{pmatrix} \,,
\end{align}
\begin{align}
    \tilde{A}^{(1)}=\frac{1}{2}\begin{pmatrix}
        0 & 0 & 0 & 0 & t^{11}_{12} p_0 & t^{12}_{12}(2-p_0) \\
        0 & 0 & 0 & t^{10}_{12} & t^{11}_{12} p_1 & t^{12}_{12}(1-p_1) \\
        0 & 0 & 0 & t^{10}_{12} & t^{11}_{12} p_1 & t^{12}_{12}(1-p_1) \\
        0 & 0 & 0 & 0 & t^{11}_{22} p_0 & t^{12}_{22}(2-p_0) \\
        0 & 0 & 0 & t^{10}_{22} & t^{11}_{22} p_1 & t^{12}_{22}(1-p_1) \\
        0 & 0 & 0 & t^{10}_{22} & t^{11}_{22} p_1 & t^{12}_{22}(1-p_1) \\
    \end{pmatrix} \,,
\end{align}
\begin{align}
    \tilde{A}^{(2)}=\frac{1}{2}\begin{pmatrix}
        0 & 0 & 0 & 0 & t^{21}_{12} p_0 & t^{22}_{12}(2-p_0) \\
        0 & 0 & 0 & t^{20}_{12} & t^{21}_{12} p_1 & t^{22}_{12}(1-p_1) \\
        0 & 0 & 0 & t^{20}_{12} & t^{21}_{12} p_1 & t^{22}_{12}(1-p_1) \\
        0 & 0 & 0 & 0 & t^{21}_{22} p_0 & t^{22}_{22}(2-p_0) \\
        0 & 0 & 0 & t^{20}_{22} & t^{21}_{22} p_1 & t^{22}_{22}(1-p_1) \\
        0 & 0 & 0 & t^{20}_{22} & t^{21}_{22} p_1 & t^{22}_{22}(1-p_1) \\
    \end{pmatrix} \,.
\end{align}
From the synchronization properties of the input and output processes and the fact that minimal presentation of the target output has only two causal states, it follows that states $1,2,3$ and $4,5,6$ of the $\tilde{\A}$ must belong to the same equivalent class and correspond to causal states $r_0$ and $r_1$ of the target output, $\A$, respectively. However, one needs to separately check the possibility for some states being \emph{transient}, that is, have long term probability of occurrence equal to 0. These states can be removed, which is equivalent to removing the corresponding row and columns of the transition matrices $\tilde{A}^{(x)}$.

We first examine the case where all states in each equivalence class are \emph{recurrent}, i.e.\@ have long term probability of occurrence greater than zero. That is, we have the two classes of states, $r_0=\{1,2,3\}$ and $r_1=\{4,5,6\}$, consisting of non-transient states. This means that the the transition matrices have a block structure with $3\times3$ sub-blocks. Note that in general two states $r$ and $r'$ are equivalent, $r\sim r'$, if the probabilities, $\Pr(w|r)=\Pr(w|r')$, assigned to every future word $w$ of length $l \in \mathbb{Z}^+$ when the machine is in state $r$ or $r'$, are  equal \cite{travers_exact_2011}. However, due to the Markov property our task of checking equivalence is greatly simplified, as we only need to compare states for equivalence up to words of length 1. Thus, for each state its future transition probabilities for each output must match the ones of other states in the same class. Under the assumption that all states are recurrent, examining the self transition of equivalence class $r_0$ with output $y=0$, we obtain $t^{01}_{11} = t^{01}_{11} = t^{02}_{11} = 0$. Similarly, from the transition from $r_1$ to $r_0$ of output $y=0$ we obtain the conditions,
\begin{align}
    t^{01}_{21} p_0 + t^{02}_{21}(2-p_0) &= t^{00}_{21} + t^{01}_{21} p_1 + t^{02}_{21}(1-p_1) \notag \\
    &=t^{00}_{21} + t^{01}_{21} p_2 + t^{02}_{21}(1-p_2) =1 \,,
\end{align}
which imply that $t^{00}_{21}=t^{01}_{21}=t^{01}_{21}=\nicefrac{1}{2}$.
Similarly, the $r_0$ to $r_0$ and $r_1$ to $r_1$ transitions of output $y=0$ imply that  $t^{10}_{12}=t^{12}_{12}=t^{22}_{12}=\nicefrac{1}{2}$ and $t^{10}_{22}=t^{11}_{22}=t^{12}_{22}=\nicefrac{1}{2}$, respectively. Finally, from output $y=2$ we find the conditions $t^{20}_{12}=t^{21}_{12}=t^{22}_{12}=\nicefrac{1}{2}$ and $t^{20}_{22}=t^{21}_{22}=t^{22}_{22}=0$. By noting the fact that in the solution $T^{(y|x)}=T^{(y|x')}$ for all $x,x'$ we see that this is an input-independent channel. In other words, it completely ignores input and always outputs the target output process $\A$ and thus corresponds to the trivial solution. It remains to examine the cases with transient states.

Even though in principle there are many different cases that one needs to consider separately, we can show that only one family of solutions exists, this time input-independent. To show this we can prove the following statements sequentially (each statement assumes the validity of the previous one):
\begin{enumerate}
    \item At least one state from each of the two classes, $r_0=\{1,2,3\}$ and $r_1=\{4,5,6\}$, needs to be recurrent. 
    \item States 5 and 6 of equivalence class $r_1$ can not be transient.
    \item State 1 of equivalence class $r_0$ can not be transient.
    \item With states 1,5,6 recurrent, states 2 or 3 of class $r_0$ alone can not be transient: they are either both recurrent or both transient.
    \item If states 2 and 3 are both transient, state 4 of equivalence class $r_1$ can't be recurrent.
\end{enumerate}
All of the above statements follow from examining the transition matrices of the resulting outputs. For example, statement 2 follows by assuming that one of the states 5 and 6 is recurrent and the other transient or both are transient, and verifying that it is impossible to find parameter values that correctly reproduce the target outputs. Specifically, the self transition from equivalence class 2 while emitting output 2 is 0 while the target output has a self transition with probability $\nicefrac{1}{2}$.

In view of the above statements only two possibilities remain:
\begin{enumerate}
    \item All states are recurrent.
    \item Only states 1,5 and 6 are recurrent.
\end{enumerate}
We have already examined the first possibility and showed that it leads to the trivial solution. It remains to explore the second case. Since, by assumption, states 2,3,4 are transient, there can be no flow of probability from the recurrent states as this would make them recurrent as well. This implies that $t^{01}_{21}=t^{02}_{21}=t^{10}_{22}=t^{20}_{22}=0$.
Moreover, since states 2,3,4 are transient  
we can remove the corresponding rows and columns from the transition matrices, which reduces them to
\begin{align}
    \tilde{A}^{(0)} = \frac{1}{2}\begin{pmatrix}
        0  & 0 & 0 \\
        t^{00}_{21} &  0 & 0 \\
        t^{00}_{21} & 0 & 0 \\
    \end{pmatrix}, \, 
    \tilde{A}^{(1)}=\frac{1}{2}\begin{pmatrix}
        0 & t^{11}_{12} p_0 & t^{12}_{12}(2-p_0) \\
        0 & t^{11}_{22} p_1 & t^{12}_{22}(1-p_1) \\
        0 & t^{11}_{22} p_1 & t^{12}_{22}(1-p_1) \\
    \end{pmatrix} \,,
\end{align}
and
\begin{align}
    \tilde{A}^{(2)}=\frac{1}{2}\begin{pmatrix}
        0 & t^{21}_{12} p_0 & t^{22}_{12}(2-p_0) \\
        0 & t^{21}_{22} p_1 & t^{22}_{22}(1-p_1) \\
        0 & t^{21}_{22} p_1 & t^{22}_{22}(1-p_1) \\
    \end{pmatrix} \,.
\end{align}
After removing states 2,3 and 4 we are left with three states. State 1 must correspond to state $r_0$ of the target while states 2 and 3 must belong to the same class and  correspond to state $r_1$ of the target. With these observations, to correctly reproduce the transitions associated with output 0, we have that $t^{00}_{21}=1$. Similarly, from the transitions from class 1 to class 2 and the self transition of class 2 with output 1 we obtain the conditions $t^{11}_{22}=t^{12}_{22}=1$ and $t^{11}_{12} p_0 + t^{12}_{12} (2-p_0)=1$. Finally, from the transition with output 2 we find $t^{21}_{22}=t^{22}_{22}=0$ and $t^{21}_{12} p_0 + t^{22}_{12} (2-p_0)=1$. From the preceding conditions and unifilarity it also follows that $t^{01}_{11}=t^{02}_{11}=0$. Thus, the general form of the 3-parameter family of \etrs{} that take $\B_3$ to $\A$ have the general form
\begin{align}
    \hat{T}:\{&\begin{pmatrix}
        t^{00}_{11} & 0 \\
        1 & 0
    \end{pmatrix}\,,
    \begin{pmatrix}
        0 & t^{10}_{12} \\
        0 & 0
    \end{pmatrix}\,,
     \begin{pmatrix}
        0 & 1-t^{00}_{11}-t^{10}_{12} \\
        0 & 0
    \end{pmatrix}\,,  \notag \\
    &\begin{pmatrix}
        0 & 0 \\
        0 & 0
    \end{pmatrix}\,,
    \begin{pmatrix}
        0 & t^{11}_{12} \\
        0 & 1
    \end{pmatrix}\,,
     \begin{pmatrix}
        0 & 1-t^{11}_{12} \\
        0 & 0
    \end{pmatrix}\,, \notag \\
    &\begin{pmatrix}
        0 & 0 \\
        0 & 0
    \end{pmatrix}\,,
    \begin{pmatrix}
        0 & 1-t^{22}_{12} \\
        0 & 1
    \end{pmatrix}\,,
     \begin{pmatrix}
        0 & t^{22}_{12} \\
        0 & 0
    \end{pmatrix}\} \,,
\end{align}
with 
\begin{align}
    t^{22}_{12} = \frac{1-(1-t^{11}_{12})p_0}{2-p_0} \,.
\end{align}
All these classical \etrs{} correctly reproduce the output $\A$ and moreover are minimal as they all have the same statistical complexity of approximately 0.918. The complexity follows from the fact that the stationary distribution of all these channels is essentially $\{\nicefrac{1}{3},\nicefrac{2}{3}\}$ for input $\B_3$, independent of the free parameters of the family of models.

The situation is different, however, when one considers the corresponding quantum models. The reason for that is that the quantum complexity will depend on the free parameters $t^{00}_{11}\,,t^{10}_{12}$ and $t^{11}_{12}$ of the classical \etrs{}. However, there is a theoretical upper bound on the overlap between the quantum causal states which is given by the maximum fidelity constraint \cite{elliott_quantum_2022}, explicitly found as
\begin{align}
    F_{12} &= \min\left\{\sqrt{t^{00}_{11}},\sqrt{t^{11}_{12}}, \sqrt{1-t^{22}_{12}}\right\} \notag \\
    &=\min\left\{\sqrt{t^{00}_{11}},\sqrt{t^{11}_{12}}, \sqrt{\frac{1-t^{11}_{12}p_0}{2-p_0}}\right\}\,.
\end{align}
The maximal minimum value of the second and third terms is simultaneously achieved for the value $t^{11}_{12}=\nicefrac{1}{2}$ irrespective of the value of $p_0$. To obtain this value, the unconstrained first term may assume any value $t^{00}_{11}\geq t^{11}_{12}$.  If we set $t^{00}_{11}=1$, it corresponds precisely to the \etr{} in the main text. With the maximal allowed overlap being $\braket{s_1}{s_2}=\nicefrac{1}{\sqrt{2}}$, we find that there can not be a quantum model with complexity less than 0.55. That is, we have shown that for any of these \etrs{}
\begin{align}
   F_{12} \leq F_{12}^{\max} = \frac{1}{\sqrt{2}}
\end{align}
which implies that the quantum complexity of the optimal model, $Q^{\text{opt}}$ must obey
\begin{align}
    Q^{\text{opt}}\geq Q^{\min} \approx 0.55
\end{align}
Note, however, that the fidelity constraint can not be saturated in general even by the optimal quantum model and thus will underestimate the optimal quantum complexity. Nevertheless, it suffices for our purposes as we just need to show that in the other direction there exists a quantum model that can go below this bound. 

In addition, in can be shown that the optimal quantum complexity has an upper bound.
Specifically, if we encode classical causal states to quantum states according to
\begin{align}
    \ket{s_1} = \left(\sqrt{t^{00}_{11}}\ket{01}+\sqrt{t^{10}_{12}}\ket{12}+\sqrt{1-t^{00}_{11}-t^{10}_{12}}\ket{22}\right) \notag \\ 
    \times \left(\sqrt{t^{11}_{12}}\ket{12}+\sqrt{1-t^{11}_{12}}\ket{22}\right) \notag \\ \times \left(\sqrt{1-t^{22}_{12}}\ket{12}+\sqrt{t^{22}_{12}}\ket{22}\right)
\end{align}
and
\begin{align}
    \ket{s_2} = \ket{01}\ket{12} \ket{12}\,.
\end{align}
their overlap is readily computed
\begin{align}
    \braket{s_1}{s_2} = \sqrt{t^{00}_{11} t^{11}_{12} (1-t^{22}_{12})} =   \sqrt{t^{00}_{11} t^{11}_{12} \frac{(1-t^{11}_{12} p_0)}{2-p_0}} \,,
\end{align}
 and attains its maximal value for $t^{00}_{11}=1$ and 
 \begin{align}
   \braket{s_1}{s_2} =  \begin{cases} 
      \sqrt{\frac{1-p_0}{2-p_0}} \,,&  t^{11}_{12}=\frac{1}{2p_0}\,, \text{when } 0<p_0\leq \nicefrac{1}{2} \\
      \frac{1}{\sqrt{4p_0(2-p_0)}} \,, & t^{11}_{12}=1\,, \,\,\,\,\,\,\text{when } \nicefrac{1}{2}<p_0\leq 1. 
   \end{cases}
\end{align} 
In either case, for values of $p_0$ close to $\nicefrac{1}{2}$, which we assume in the main text, we find that the overlap is close to $\nicefrac{1}{\sqrt{3}}$ and thus the quantum model can reach a quantum complexity of approximately 0.682. We have thus obtained the following inequality that upper and lower bounds the optimal quantum complexity,
\begin{align}
    0.682 \geq Q^{\text{opt}}_{\B_3\rightarrow \A}\geq 0.55\,.
\end{align}
It follows that the quantum complexity of the optimal quantum model is strictly less than the classical complexity but it can never reach a value less than about 0.55.

So far, we have only discussed the case with $n=3$ but the proof generalises in a straight forward manner in the case with $n>3$. Specifically, first it can be shown that no memoryless channel that takes $\B_n$ to $\A$ exists, that is, the minimal \etr{} has to have at least two causal states and is Markovian. Then, one can evaluate the output and the states will split in two classes $r_1=\{1,\ldots,n\}$ and $r_2=\{n+1, \ldots, 2n\}$ that correspond to the two causal states of the output. As with the case of $n=3$ it can be shown that 
\begin{enumerate}
    \item At least one state from each of the two classes, $r_1=\{1,\ldots,n\}$ and $r_2=\{n+1,\ldots,2n\}$, needs to be recurrent. 
    \item States $n+2,\ldots, 2n$ of equivalence class $r_2$ can not be transient.
    \item State 1 of equivalence class $r_1$ can not be transient.
    \item With states $1,5,\ldots,2n$ recurrent, states $2,\dots, n$ of class $r_1$ alone can not be transient: they are either all recurrent or all transient.
    \item If states $2,\dots, n$ are all transient, state $n+1$ of equivalence class $r_2$ can't be recurrent.
\end{enumerate}
which once again leads to two possibilities:
\begin{enumerate}
    \item All states are recurrent.
    \item Only states $1,n+2,\ldots,n$ are recurrent.
\end{enumerate}
The first always corresponds to the trivial solution while the second leads to a family of input-dependent solutions, all of which have a classical complexity of 0.918 when driven by $\B_n$ as in the case with $n=3$. From the structure of the obtained solutions, the maximal fidelity constraint is found as 
\begin{align}
    F_{12} 
    =\min&\left\{\sqrt{t^{00}_{11}},\sqrt{t^{11}_{12}}, \sqrt{t^{12}_{12}} , \ldots , \sqrt{t^{1, n-2}_{12}}\,, \right. \notag \\
    &\,\,\,\,\,\,\,\,\, \left.\sqrt{\frac{1-t^{11}_{12}p_{01}-\ldots-t^{1,n-2}_{12}p_{n-2}}{2-p_{01}-\ldots-p_{n-2}}}\right\}\,.
\end{align}
The maximal minimum value is attained when $t_{12}^{11}=\ldots=t_{1, n-2}^{11}=\nicefrac{1}{2}$ and the unconstrained first term obeys $t_{12}^{11}\geq \nicefrac{1}{2}$. Once again, if we set $t_{12}^{11}=1$ we obtain the \etr{} described in the main text.

We can also derive an upper bound on the quantum complexity, which is between the value $Q_{\B_3\rightarrow \A}\approx0.833$ for the case $n=4$ and approaches the classical complexity from below for increasing $n$. To see that, consider the \etrs{} of the main text and the encoding of causal states of the classical \etrs{} to the quantum states as $\ket{s_i}=\otimes_x\ket{s_i^x}$ with $\ket{s_i^x}=\sum_{k}\sum_{y} \sqrt{T_{ik}^{(y|x)}} \ket{y}\ket{k}$~\cite{elliott_quantum_2022,thompson_causal_2018}, from which we explicitly obtain
\begin{align}
	\ket{s_0^x}&=\frac{1}{\sqrt{2}}\ket{1}\left(\ket{1}+\ket{2}\right)  \,, \notag \\
	\ket{s_1^x}&=\ket{1}\ket{1}  
\end{align}
for all $x\neq0$, while
\begin{align}
	\ket{s_0^0}=\ket{s_1^0}=\ket{0}\ket{0}\,,
\end{align}
when $x=0$.
It follows that
\begin{align}
	\braket{s_1}{s_2} = \left(\frac{1}{\sqrt{2}}\right)^{n-1}.
\end{align}
The Gram matrix is:
\begin{align}
	G=
	\begin{pmatrix}
		\frac{1}{3} & \frac{\sqrt{2}}{3} \left(\frac{1}{\sqrt{2}}\right)^{n-1}  \\
		\frac{\sqrt{2}}{3}  \left(\frac{1}{\sqrt{2}}\right)^{n-1}  & \frac{2}{3} \\
	\end{pmatrix} \,,
\end{align}
and its eigenvalues are
\begin{align}
	\lambda_{\pm} = \frac{1}{2} \pm \frac{\sqrt{1+2^{-n+4}}}{6}.
\end{align}
The quantum entropy is then
\begin{align}
	Q_{\B_n\rightarrow\A}^{\text{upper}} = - \sum_{i=\pm} \lambda_i \log \lambda_i \,.
\end{align}
For $n=4$ we obtain the value $Q_{\B_3\rightarrow \A}^{\text{upper}}\approx0.833$. The value increases with increasing $n$ and in the limit of large $n$ we recover the classical distribution $\hat{\varphi} = \{\nicefrac{1}{3},\nicefrac{2}{3}\}$ as expected.
Thus the quantum complexity of an optimal model will assume a value constrained by the double inequality
\begin{align}
    Q_{\B_n\rightarrow \A}^{\text{upper}} \geq Q^{\text{opt}}_{\B_n\rightarrow \A}\geq 0.55\,.
\end{align}

The process to obtain the channels from $\A$ to $\B_n$ is in principle the same but the number of classical solutions is in general bigger than that of the case from $\B_n$ to $\A$, which makes a systematic presentation not practical. However, for our purposes it suffices to show that there do not exist \etrs{} that can beat the classical complexity of producing $\B_n$ regardless of input. We first note that in process $\A$ all probabilities are \nicefrac{1}{2}, while, on the other hand, by assumption all emissions of $\B_n$ happen with probabilities $p_{i,j}\neq p_{k,l} \neq \nicefrac{1}{2}$ for all $i,j,k,l$. With increasing $n$ it becomes obvious that more states are necessary to correctly reproduce the output process $\B_n$. In fact, one needs at least the same number of states, $n$, as that of $B_n$, otherwise the probabilities can not be reproduced. It remains to show that the distribution over the causal states of the \etrs{} and of $B_n$ do in fact coincide. Since the \etrs{} that map $\A$ to $\B_n$ are of pure feedback Markov order 1, the transition matrices of the joint process $J^{(x,y)}$ are split into sub-blocks, each of which, corresponds to a unique state of the output process $B_n$, as well as a unique state of the \etrs{}. Since the flow of probability in and out of these sub-blocks follows that of the output process, they must have the same stationary distribution and thus the same statistical complexity. Thus, all these maps when driven by $\A$ have the same statistical complexity as that of the process $\B_n$.  

\section{Appendix B: Classical complexity of $\hat{T}_n$ }
First we obtain the stationary distribution of process $\A$ by solving the simultaneous equations $\pi_0=\nicefrac{\pi_1}{2}$ and $\pi_0+\pi_1 =1$ 
the first of which follows directly from Fig.\@ \ref{Fig: processes A and B} and the second reflects the fact that $\pi$ is a probability distribution. The solution is $\pi = (\pi_0, \pi_1) =\left(\nicefrac{1}{3},\nicefrac{2}{3}\right)\,. $
Having obtained the stationary distribution, we can calculate the probabilities of emitting a certain symbol through $\Pr(x)=\sum_{i=0,1} \pi_i \Pr(x|\pi_i)\,.$
We explicitly find the values $\left\{\Pr(0),\Pr(1),\Pr(2)\right\}=\left\{\nicefrac{1}{3},\nicefrac{1}{2},\nicefrac{1}{6}\right\}$. Then, by using the properties of the \etrs{} $\T_n$, we can evaluate the stationary distribution given the input $\A$. Specifically, by noting that every output of the \etrs{} leads to a transition to the state with the same label, the equations for the stationary distribution $\varphi$ are given in terms of the probabilities of receiving a symbol 0,1 or 2 from the input, $\Pr(0),\Pr(1)$ and $\Pr(2)$.
For $n\geq3$ and under the assumption $p_{i,j}\approx \nicefrac{1}{n-1}\,, \forall i,j$ we have the stationary distribution
\begin{align}
	\varphi_0 &= \Pr(0)\,\,,\,\,\varphi_2 = \frac{\Pr(1)}{n-1} +\Pr(2)\,\,, \notag \\
	\varphi_j &= \frac{\Pr(1)}{n-1} \,,  j=1,3,\ldots,n-1 \,.
\end{align}
The classical complexity follows
\begin{align}
	C_{\A\rightarrow \B_n} =& -\Pr(0)\log\Pr(0) \notag \\
 -&(n-2)\left(\frac{\Pr(1)}{n-1}\right)  \log\left(\left(\frac{\Pr(1)}{n-1}\right)\right) \notag \\
	-&\left( \frac{\Pr(1)}{n-1} +\Pr(2)\right) \log\left( \frac{\Pr(1)}{n-1} +\Pr(2)\right) .
\end{align}

\section{Appendix C: The upper bound $Q^{\text{upper}}_{\A \rightarrow \B_n}$.}
We can derive an explicit upper bound, $Q^{\text{upper}}_{\A \rightarrow \B_n}$, on the quantum entropy of the \etr{} from $\A$ to $\B_n$ in terms of a small perturbation parameter $\delta$. 

Note that in the limit where all probabilities are the same, i.e. $p_{ij}=p_{kl}=\nicefrac{1}{n-1}\,, \forall i,j,k,l$, the memory state becomes pure since $\rho=\sum_i \varphi_i \ketbra{s_i}{s_i} = \ketbra{s}{s}$, as all quantum causal states reduce to the same state $\ket{s_i}\equiv\ket{s}=\ket{0}\ket{0}\left(\sum_{j=1}^n \frac{\ket{j}\ket{j}}{\sqrt{n-1}}\right)\ket{2}\ket{2}\ldots \ket{n}\ket{n}$ for all $i$. 

On the other hand, if we make a small perturbation of the probabilities around the value $\nicefrac{1}{n-1}$ so that $p_{ij}=\frac{1+\delta_{ij}}{n-1}$, with $\abs{\delta_{ij}}\leq \delta$ for some small $\delta>0$, we have the state of the memory $\sigma = \sum_i \varphi_i \ketbra{s_i}{s_i}$. Then, the fidelity of the perturbed state, $\sigma$, with the memory state at the limit, $\rho$, is lower bounded according to
\begin{align}
    F(\rho,\sigma)&=\bra{s}\sigma\ket{s} = \sum_i \varphi_i \left(\sum_{j=1}^{n-1} \sqrt{\frac{p_{ij}}{n-1}}\right)^2 \notag \\
    &\geq  \frac{1}{2} + \frac{1}{2}\sqrt{1-\delta^2} \equiv F_{\text{min}}(\delta) \,,
\end{align}
where we have used the fact that the minimum of a concave function over a convex set must be attained at the boundary. Specifically, the minimum occurs when half the $p_{ij}$ are equal to $\frac{1+\delta}{n-1}$ and the other half to $\frac{1-\delta}{n-1}$ for the even case, while for the odd case one is equal to $\frac{1}{n-1}$.
This immediately implies an upper bound on the trace distance between the two states, $T \equiv T(\rho,\sigma) = \nicefrac{\norm{\rho-\sigma}_{1}}{2}$, through the inequality $T(\rho,\sigma)\leq \sqrt{1-F(\rho,\sigma)}$, that is,
\begin{align}
    T(\rho,\sigma)\leq T_{\max} (\delta) = \sqrt{1-F_{\min} (\delta)}\,.
\end{align}
Up to third order in $\delta$ we see that $T_{\max} (\delta)=\frac{\delta}{2}+\mathcal{O}(\delta^3)$.

Finally we employ a result that upper bounds the difference of von Neumann entropies in terms of their trace norm distance \cite{audenaert_sharp_2007}. Specifically,
\begin{align}
    \abs{S(\rho)-S(\sigma)} \leq T \log_2(d-1) + H({T,1-T}) \,,
\end{align}
where $T \equiv T(\rho,\sigma)$ denotes the trace norm distance and $H$ denotes the Shannon entropy, $H(p)=-\sum_i p_i \log_2 p_i$. Noting that the function on the right hand side of last inequality is monotonically increasing in the range $T\in [0,\nicefrac{d-1}{d}]$ and the fact that for sufficiently small $\delta$ and large $n$, $T_{\max}(\delta)$ is always inside this range, an upper bound is obtained by substituting $T_{\max} (\delta)$ for $T$.   

Since the von Neumann entropy of $\rho=\ketbra{s
}{s}$ is zero as it is a pure state, we have the upper bound on the quantum entropy of $\sigma$, $S(\sigma)\equiv Q_{\A \rightarrow \B_n} \leq Q^{\text{upper}}_{\A \rightarrow \B_n}$,
\begin{align}
    Q^{\text{upper}}_{\A \rightarrow \B_n}
    = T_{\max} (\delta) \log_2 (n-1) + H(T_{\max} (\delta),1-T_{\max} (\delta))\,.
\end{align}
For example, in the case of $n=3$ and for perturbations no larger than $\delta=10^{-2}$ the quantum complexity is upper bounded as $Q=S(\sigma)\leq 0.097$. In practice, however, the quantum complexity can be significantly lower than the bound predicts. Nevertheless, it suffices for our purposes.

Note that even though the causal states as defined in the main text are embedded in a Hilbert space of dimension $(\abs\Y\abs\S)^{\abs\X}$, there exists a unitary transformation that maps them to a subspace of dimension of at most $\abs\S$ \cite{thompson_using_2017}. Thus, we can set $d=n$ and the last bound follows. \

The above inequality bound should be interpreted in the following way: given a value of $n$ there exists a choice of $\delta$ that upper and lower bounds the perturbation of the probabilities of the family of processes $\B_n$ around the value $\frac{1}{n-1}$ such that the quantum complexities of the models for all values of $n^\prime \leq n$ are sufficiently close to 0. This in turn shows, that for every value of $n$ there exist processes $\A$ and $\B_n$ such that the channels from $\A$ to $\B_n$ have classical complexities logarithmically diverging with $n$ while their quantum complexities are sufficiently close to 0, and certainly less than the minimum theoretical bound for the quantum complexity of the channels from $\B_n$ to $\A$.

\end{document}